\documentstyle[preprint,aps,epsf]{revtex}

\begin{document}
\draft

\title{
Enhanced Pulse Propagation
in Non-Linear Arrays of Oscillators
}
\author{Antonio Sarmiento\footnote{Permanent address:
Instituto de Astronom\'{\i}a, Apdo. Postal 70-264, Ciudad Universitaria,
M\'{e}xico D. F. 04510, M\'{e}xico},
Ramon Reigada\footnote{Permanent address:
Departament de Qu\'{\i}mica-F\'{\i}sica, Universitat de Barcelona,
Avda. Diagonal 647, 08028 Barcelona, Spain},
Aldo H. Romero\footnote{Present address: Max-Planck Institut f\"{u}r
Festk\"{o}rperforschung, Heisenbergstr. 1, 70569 Stuttgart, Germany},
and Katja Lindenberg}

\address
{Department of Chemistry and Biochemistry 0340\\
University of California San Diego\\
La Jolla, California 92093-0340}

\date{\today }
\maketitle

\begin{abstract}
The propagation of a pulse in a nonlinear array of oscillators is
influenced by the nature of the array and by its coupling
to a thermal environment.  For example, in some arrays a pulse
can be speeded up  while in others a pulse can
be slowed down by raising the temperature.  We begin by showing that
an energy pulse (1D) or energy front (2D) travels more rapidly and
remains more localized over greater distances in an isolated array
(microcanonical) of hard springs than in a harmonic array
ot in a soft-springed array.
Increasing the pulse amplitude causes it to speed up in
a hard chain, leaves the pulse speed unchanged in a harmonic system,
and slows down the pulse in a soft chain.
Connection of each site to a thermal environment
(canonical) affects these results very differently
in each type of array. In a hard chain the dissipative forces slow down
the pulse while raising the temperature speeds it up.
In a soft chain the opposite occurs: the dissipative forces actually
speed up the pulse while raising the temperature slows it down.
In a harmonic chain neither dissipation nor temperature changes affect
the pulse speed. These and other results are explained on the basis of
the frequency vs energy relations in the various arrays.

\end{abstract}


\section{Introduction}
\label{sec:level1}

In recent years there has been a great deal of interest in
the interplay of nonlinearity and applied forcing (deterministic and/or
stochastic) in the stationary and transport properties of discrete
spatially extended
systems \cite{general1}.
The ability of discrete anharmonic arrays to localize and
propagate energy in a persistent fashion, and the fact that noise
may act (sometimes against one's intuition) to enhance these
properties, has led to particularly intense
activity \cite{general2,temp,sanchobook}. Interesting noise-induced
phenomena include stochastic resonance \cite{sr}, noise-induced phase
transitions \cite{chris}, noise-induced front propagation \cite{sancho},
and array-enhanced stochastic resonance \cite{lindner}.

Our interests in this area have been motivated by the relative
dearth of information concerning the effects of a thermal
environment on the sometimes exquisite balances that are required
to achieve these interesting resonances and persistences
\cite{temp,ourchain1,ourchain3}. At the same
time, we have also noted that most of the literature has
concentrated on overdamped arrays (often motivated by mathematical
or computational constraints rather than physical considerations),
a restriction that leaves out important inertial effects and that
is easily overcome.

Perhaps the simplest generic discrete arrays in which to analyze
these issues are systems of oscillators consisting of masses that
may be subject to  local monostable potentials (harmonic or
anharmonic) and nearest neighbor  monostable
interactions (harmonic or anharmonic) (other generic arrays of
current interest are bistable units linearly or nonlinearly
connected to one another).  These are the systems of choice in our
work, and we have separated our inquiries into three distinct
groups of questions: 1) The study of such arrays in thermal
equilibrium \cite{ourchain1}.  The questions here concern the
spatial and temporal ``energy landscape" that determines the
degree of spontaneous energy localization due to thermal
fluctuations and the temporal persistence of high or low energy
regions; 2) The study of the propagation of a persistent signal
applied at one end of the array \cite{ourchain3}.  The questions
here concern the signal-to-noise ratio and distance of signal
propagation; 3) The study of the propagation of an initial
$\delta$-function energy pulse (this work).  The questions here
concern the velocity of propagation and the dispersion of such a
pulse.

It is useful and relevant to provide a very brief summary of our
conclusions on the first two sets of questions.
Our work on equilibrium energy landscapes \cite{ourchain1} was
based on  chains  of harmonically coupled oscillators subject to a
local potential that may be anharmonic.
Each oscillator is connected to a heat bath
at temperature $T$.
We analyzed the thermal fluctuations and their
persistence  as influenced by the local potential
(we compared hard, harmonic, and soft potentials),
the strength of the harmonic coupling between the oscillators, the
strength of the dissipative force connecting each mass to the
heat bath, and the temperature. Among our
conclusions are the following: 1) An increase in
temperature in weakly coupled soft chains leads
not only to greater energy fluctuations but also to a
slower decay of these fluctuations;
2) An increase in temperature in weakly dissipative hard chains leads
not only to greater energy fluctuations but also to a slower decay of
these fluctuations; 3) High-energy-fluctuation mobility in
harmonically coupled nonlinear chains in thermal equilibrium does {\em
not} occur beyond that which is observed in a completely harmonic chain.

However, we noted earlier that interest in energy
localization in perfect arrays, as contrasted with localization induced by
disorder, arises in part because localized energy in these systems may be
{\em mobile}.  Dispersionless or very slowly dispersive mobility
would make it possible for localized energy to reach a predetermined
location where it can participate in a physical or chemical event.
Our results raised the possibility of observing
such localized mobility if the anharmonicity lies in the interoscillator
interactions rather than (or in addition to) the local potentials.
We ascertained that a persistent sinusoidal force applied to one site of
a chain of masses connected by anharmonic springs
may indeed propagate along the chain \cite{ourchain3}.
Furthermore, we demonstrated a set of resonance phenomena that we have
called  {\em thermal resonances} because they involve optimization via
{\em temperature} control.  In particular, these results establish
the existence of
optimal finite temperatures for the enhancement of the signal-to-noise
ratio at any site along the chain, and of an
optimal temperature for
maximal distance of propagation along the chain.
These resonances differ from the usual
noise-enhanced propagation where the noise is external and/or the system is
overdamped.

This work  addresses the third set of questions posed above
concerning the way in which a nonequilibrium initial condition
in the form of an energy pulse propagates as the system relaxes toward
equilibrium. More specifically, we investigate the motion and
dispersion of such an energy pulse and the
effects of finite temperatures on pulse propagation.  In view of
our earlier results on thermal resonances, perhaps the most
interesting question to be asked at this point is this: Is it
possible to enhance pulse propagation  via temperature control?

In order to monitor the evolution of the nonequilibrium initial condition
it is useful to partition the Hamiltonian as
\begin{equation}
H=\sum_n E_n
\end{equation}
where $E_n$ contains the kinetic energy of site $n$ and an appropriate
portion of the potential energy of interaction with its nearest neighbors
(1/2 in one dimension, 1/4 in two dimensions).  In one dimension
\begin{equation}
E_n = \frac{p_n^2}{2} +\frac{1}{2} V(x_{n+1},x_n)+
\frac{1}{2}V(x_n,x_{n-1})~.
\label{localenergy}
\end{equation}

In Section~\ref{pots} the potentials considered in
this paper are briefly presented.  Section~\ref{oned} contains
our analysis and results for one-dimensional
oscillator chains.  Here we discuss ways to
characterize the mobility and dispersion of an initial localized
impulse, and compare the behaviors of harmonic, hard anharmonic, and
soft anharmonic chains.  In
Section~\ref{twod} we present some results for isolated
two-dimensional arrays and note some interesting geometric
features with perhaps unanticipated consequences.
Section~\ref{conclusions} is a summary of results.

\section{Potentials}
\label{pots}

\begin{figure}[htb]
\begin{center}
\leavevmode
\epsfxsize = 3.0in
\epsffile{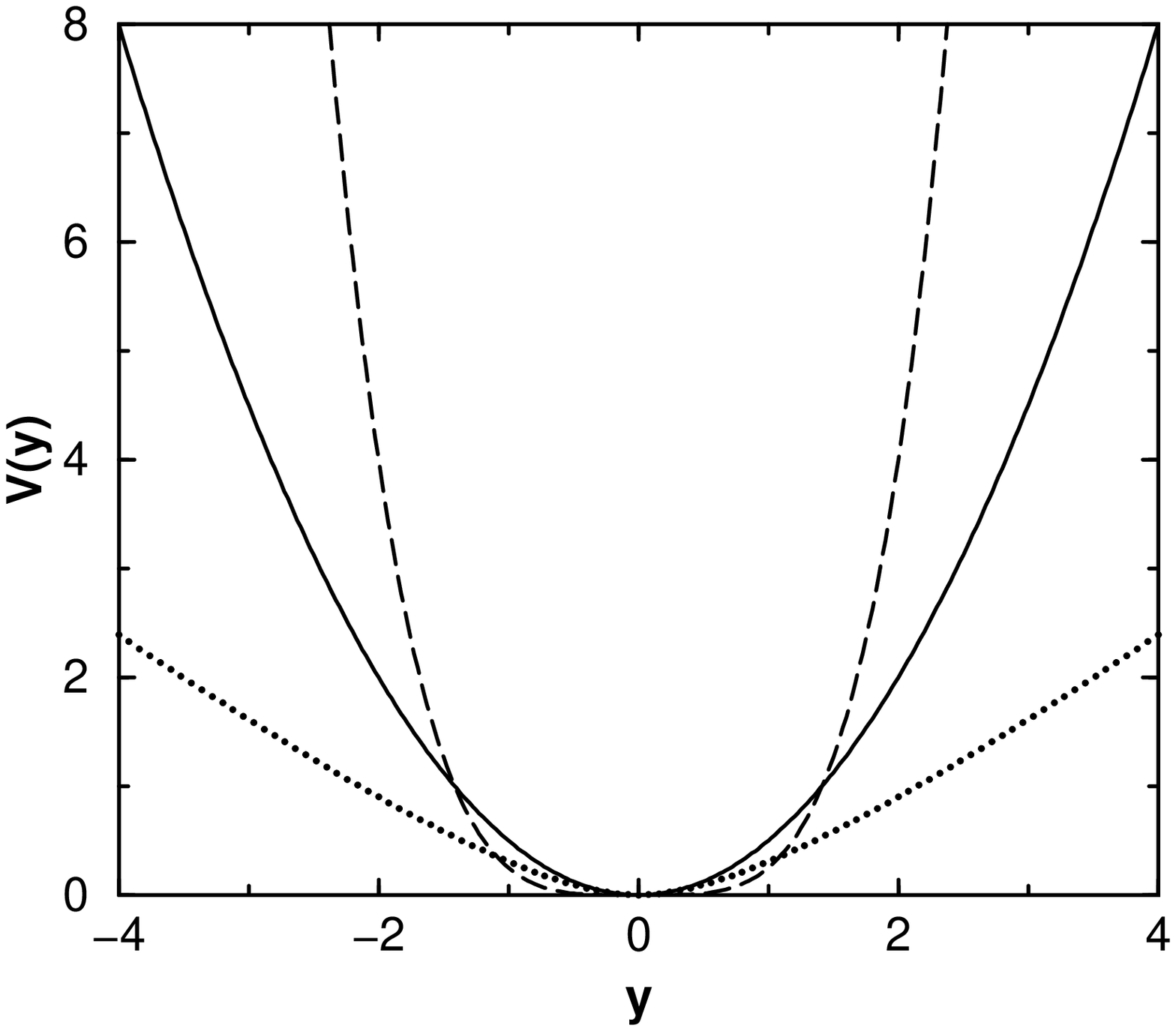}
\leavevmode
\epsfxsize = 3.0in
\epsffile{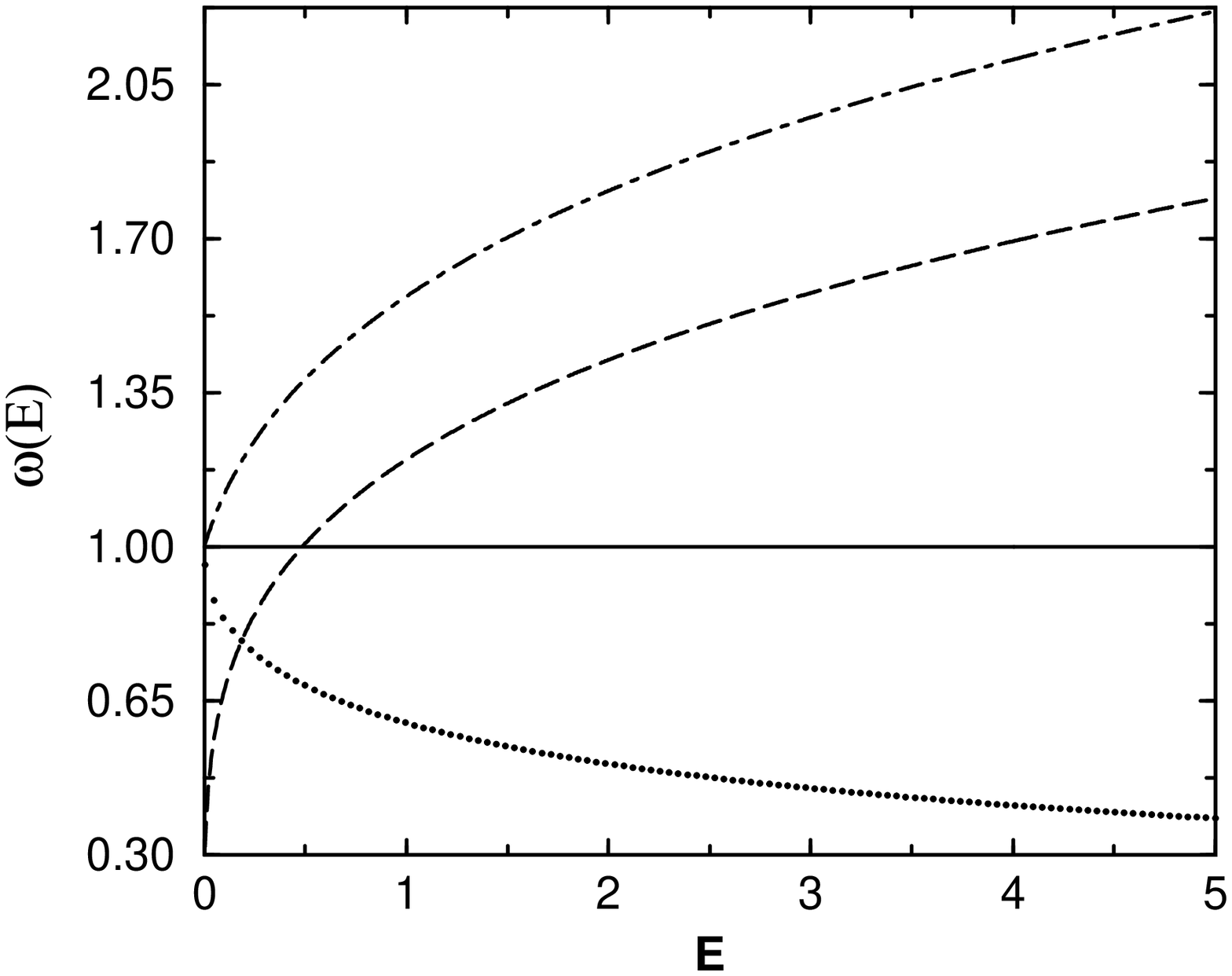}
\vspace{-0.2in}
\end{center}
\caption
{First panel: the potentials defined in
Eqs. (\ref{harmonicpot})-(\ref{softpot}) with $k=1$.
Solid curve : harmonic potential, $V_0(y)$.
Dashed curve: hard anharmonic potential, $V_h(y)$.
Dotted curve: soft anharmonic potential, $V_s(y)$.
Second panel: frequency
as a function of the oscillator energy for these potentials.
The dot-dashed line shows the frequency curve for the commonly
used potential $V_0(y)+V_h(y)$.}
\label{fig1}
\end{figure}

The particular potentials as a function of the relative
displacement $y\equiv x_n-x_{n-1}$ used in our presentations
are the harmonic,
\begin{equation}
V_0(y)= \frac{k}{2} y^2,
\label{harmonicpot}
\end{equation}
a hard anharmonic,
\begin{equation}
V_h(y)= \frac{k}{4} y^4,
\label{hardpot}
\end{equation}
and a soft anharmonic,
\begin{equation}
V_s(y) = k\left[|y| -\ln (1+|y|)\right]~.
\label{softpot}
\end{equation}
The anharmonic potentials have
been chosen to be strictly hardening and strictly softening, respectively,
with increasing amplitude.  The potentials are shown in the first panel
in Fig.~\ref{fig1}.  In almost all our simulations we take $k=1$.


The displacement variable $y$ of a single oscillator
of energy $E$ in a potential $V(y)$ satisfies the equation of motion
\begin{equation}
\frac{dy}{dt} = \pm \sqrt{2[E-V(y)]}~.
\end{equation}
This equation can be integrated and, in particular, one can express the
period of oscillation $\tau(E)$ and the frequency of oscillation $\omega(E)$
as
\begin{equation}
\tau(E) = \frac{2\pi}{\omega(E)} = 4\int_0^{y_{max}} \frac{dy}{\sqrt
{2[E-V(y)]}}~.
\label{frequency}
\end{equation}
The amplitude $y_{max}$ is the positive solution of the
equation $V(y)=E$. The resulting oscillation frequencies obtained from
the integration of Eq.~(\ref{frequency}) for the three
potentials with $k=1$ as well as that of the frequently
used ``quadratic plus quartic potential"
are shown in the second panel of Fig.~\ref{fig1}~\cite{pippard}.

The frequency vs energy variations seen in Fig.~\ref{fig1} can be shown
via rescaling and bounding arguments to represent general features of
hardening and softening monostable potentials.  The exercise is trivial if
the potential is of the form
\begin{equation}
V(y)=\frac{k}{\alpha} y^{\alpha}
\label{onepower}
\end{equation}
since then
\begin{eqnarray}
\tau(E) &=& 4\int_0^{y_{max}}
\frac{dy}{\sqrt{2[E-ky^{\alpha}/\alpha]}}\nonumber\\ \nonumber\\ &
=& 4 \left(\frac{\alpha}{k}\right)^{1/\alpha} \int_0^1 \frac{dz}
{\sqrt{2(1-z^\alpha)}} ~ E^{\frac{1}{\alpha}-\frac{1}{2}}~ \equiv~
B_\alpha E^{\frac{1}{\alpha}-\frac{1}{2}}
\label{oneintegral}
\end{eqnarray}
whence
\begin{equation}
\omega(E) = \frac{2\pi}{B_\alpha}
E^{\frac{1}{2}-\frac{1}{\alpha}}~.
\label{omega}
\end{equation}
The coefficient $B_\alpha$ can be expressed exactly in terms of the beta
function and is equal to $2\pi$ for the harmonic potential.

If the potential is not of the simple single-power form it is
still possible to bound the resulting energy dependence to establish
the trend~\cite{pippard}.  For example, the soft potential
(\ref{softpot}) is bounded below by $(k/2)|y|$ and above by
$k|y|$.  These bounds immediately lead to the conclusion that the
associated $\omega(E)$ must decrease as $E^{-1/2}$. The argument
for a mixed power potential such as $V(y)=\frac{1}{2}y^2 +\frac{1}{4}y^4$
is a bit more cumbersome
but otherwise similar: by making the change of variables from $y$
to $\frac{1}{2}y^2 +\frac{1}{4}y^4 = Ez^4$ one can show not only
that $\omega(E)$ is an increasing function of $E$ but that it lies
above the harmonic potential result for any positive $E$.

Figure~\ref{fig1} summarizes the well known frequency characteristics
of oscillators: for a harmonic oscillator the frequency is independent of
energy (and, with our parameters, equal to unity); for a hard oscillator
the frequency {\em increases} with energy,
while that of a soft oscillator {\em decreases} with energy.  The hard
oscillator frequency curve starts below the other two if a harmonic
portion is not included.  These frequency--energy trends are
generalized to oscillator chains in Appendix~\ref{a}.
The frequency vs energy behavior will figure prominently in our
subsequent interpretations.  In particular, the following broad view
seems to be overarchingly supported: the speed and dispersion of
pulse propagation in discrete arrays of oscillators
are principally dependent on
the mean frequency associated with the energy in the pulse.
Higher frequencies lead to faster propagation and slower
dispersion.

\section{One-Dimensional Arrays}
\label{oned}

We consider one-dimensional arrays of $2N+1$ sites numbered from $-N$ to
$N$ with periodic boundary conditions.
We distinguish isolated chains (that is, ones not
connected to a heat bath), chains connected to a heat bath at zero
temperature, and finite temperature chains.  This provides an opportunity
to organize the effects of different parameters on the
behavior of the chains.

In all cases at time $t=0$ a kinetic energy $\varepsilon$ is imparted to one
particular oscillator (the oscillator at $n=0$) of the chain.  If the chain
is isolated or at zero temperature, this initial impulse is applied to an
otherwise quiescent chain.  At finite temperatures the chain is first
allowed
to equilibrate and then this impulse is imparted in addition to the thermal
motions already present.  We then observe how this initial $\delta$-function
impulse propagates and spreads along the chain, and how these behaviors
depend on system parameters.

\subsection{Isolated Chains}
\label{isolated}

The equations of motion for an isolated chain are
\begin{equation}
\ddot{x_n} = -\frac{\partial}{\partial x_n} [V(x_n -x_{n-1}) +
V(x_{n+1}-x_n)] ~.
\label{langn}
\end{equation}
The initial conditions are
\begin{eqnarray}
&&x_n(0)=0 \qquad{\rm for~all}~n, \nonumber\\
&&\dot{x}_n(0)=0 \qquad {\rm for}~n\neq 0, \nonumber\\
&&\dot{x}_0(0)\equiv p_0 =\sqrt{2\varepsilon}~.
\label{initialc}
\end{eqnarray}
For a harmonic array this system can of course be solved exactly, and
we do so in Appendix~\ref{b}.  The analytic harmonic results are
helpful and informative,
although our discussion is primarily based on simulation results since the
anharmonic chains cannot be solved analytically.
The numerical integration of the equations of motion
is performed using the second order Heun's method
(which is equivalent to a second order Runge Kutta
integration) \cite{Gard,Toral}  with a time step $\Delta t = 0.0001$.

One can think of the dynamics ensuing from the initial momentum
impulse
in two equivalent ways.  One is to interpret the $x_n$ and $\dot{x}_n$
as displacements and momenta
{\em along} the chain.  Two symmetric
pulses start from site zero and move to the left and to the right along
the chain, and our discussion focuses on either of these two
identical pulses.  This symmetry occurs regardless of the
{\em sign} of the initial momentum since the energy does not depend
on the sign, i.~e., the contraction of the spring between sites $n=0$ and
$n=1$ that follows an initial positive impulse has exactly the same effect as the
equal extension of the spring between sites $n=0$ and $n=-1$.
Alternatively, one can think of $x_n$ and $\dot{x}_n$ as
displacements and momenta perpendicular to the chain, the sign then
simply representing motion ``up" or ``down".  The symmetry
around the site $n=0$ is then even more obvious.

In any case, the energy $\varepsilon$ excites
the displacements as well as momenta of other oscillators as it
moves and disperses. The evolution can be characterized
in a number of ways.  We have found the most useful to be the mean
distance of the pulse from the initial site, defined as
\begin{equation}
\left< x\right> \equiv
\frac {\sum_n |n|E_n} {\sum_n E_n}
\end{equation}
and the dispersion
\begin{equation}
\sigma^2 \equiv \left< x^2 \right> -\left< x \right>^2 =
\frac {\sum_n n^2E_n} {\sum_n E_n} - \left< x \right>^2~.
\end{equation}
(The sums over $n$ extend from $-N$ to $N$.)
Here the $E_n$ are the local energies defined in Eq.~(\ref{localenergy})
and, since these depend on time, so do the mean distance and the variance.
The time dependence of the mean distance traveled is a measure of
the velocity of the pulse, and that of the dispersion is a
measure of how long the pulse survives before it degrades to a
uniform distribution.  An indication of
the progression of a pulse is shown in Appendix~\ref{b} for a harmonic
chain.

\begin{figure}[htb]
\begin{center}
\leavevmode
\epsfxsize = 3.0in
\epsffile{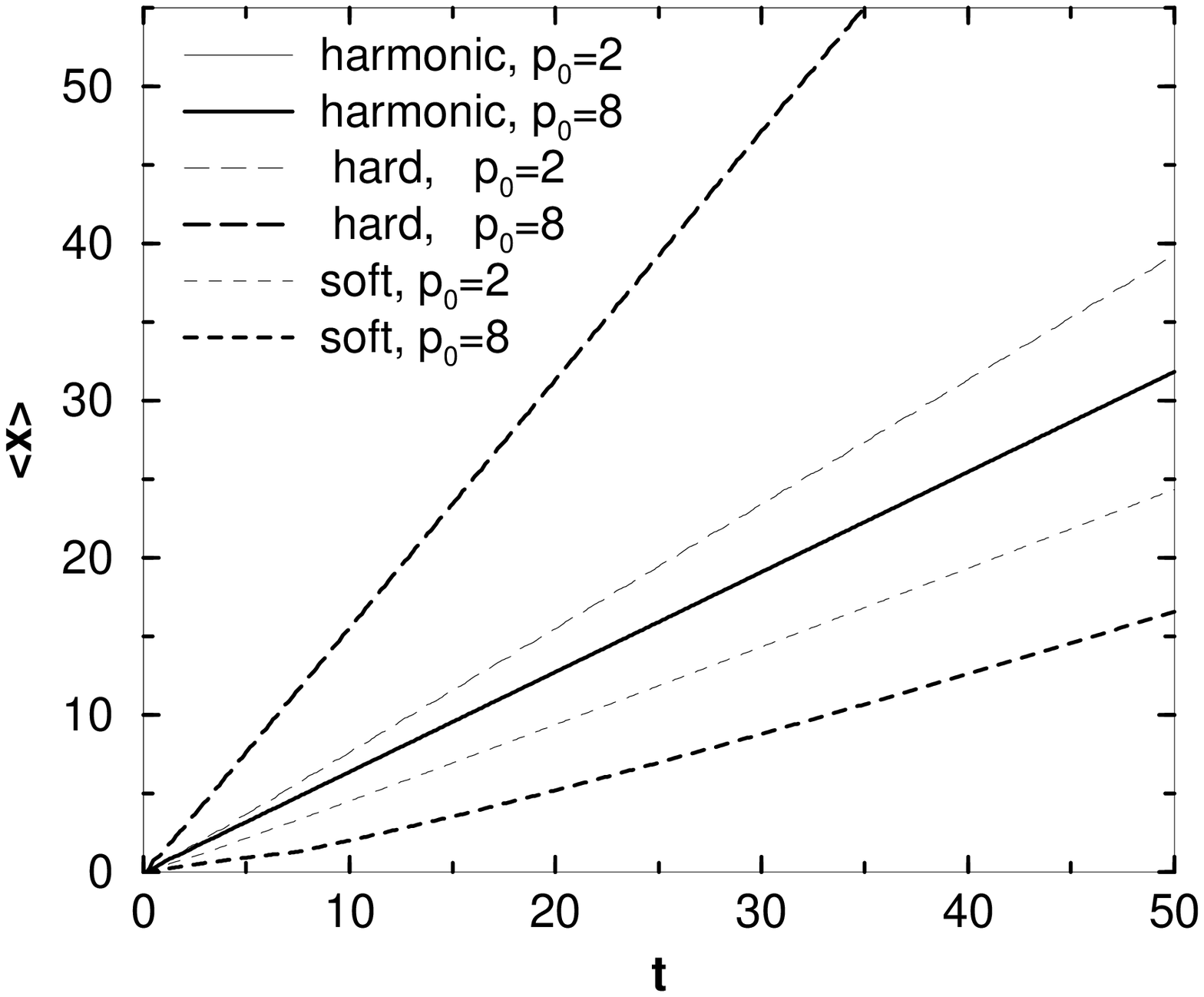}
\leavevmode
\epsfxsize = 3.0in
\epsffile{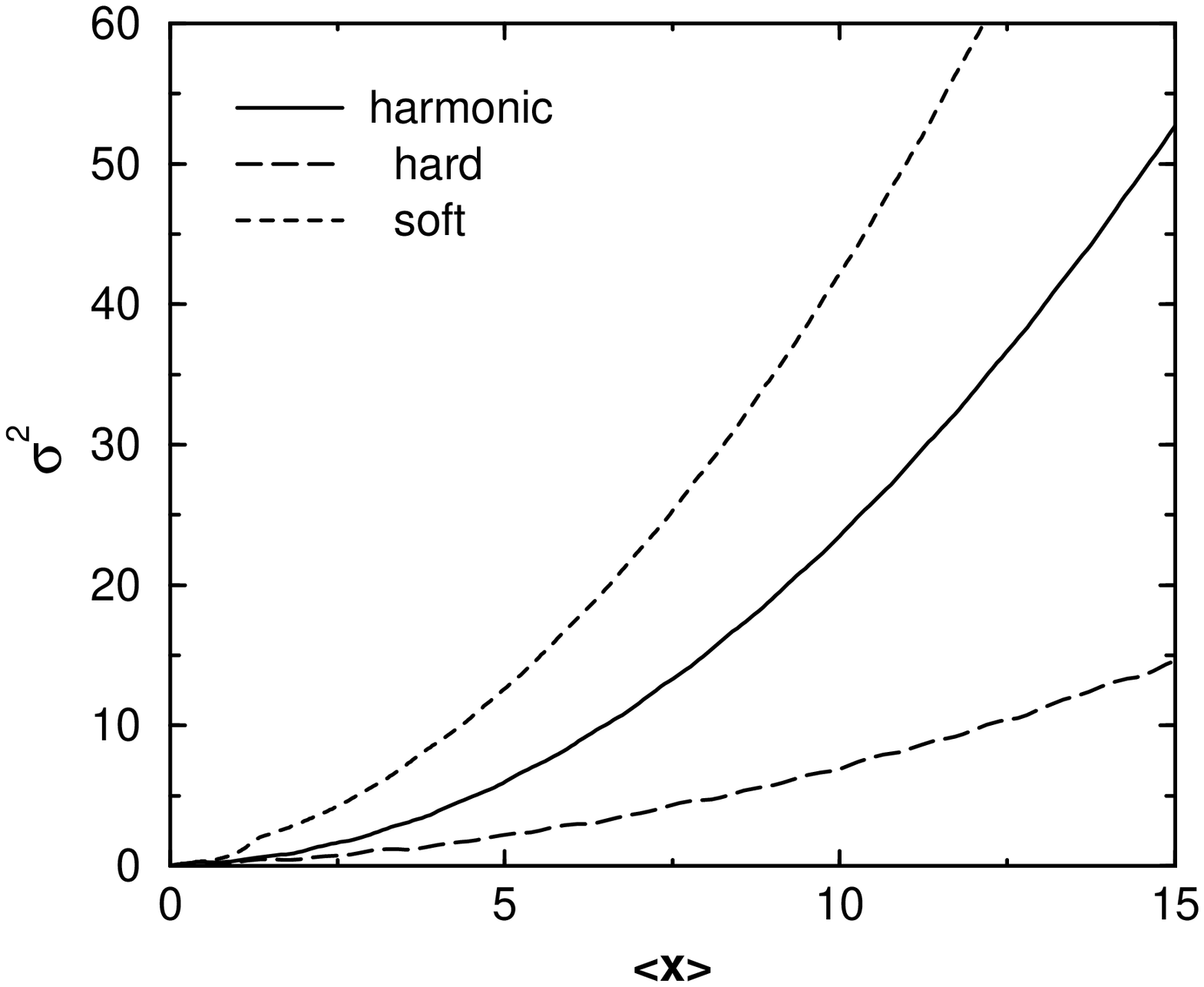}
\vspace{-0.2in}
\end{center}
\caption
{
First panel: Mean distance traveled by the initial energy pulse
as a function of time for the
hard, harmonic, and soft chains with several initial
momentum
amplitudes.  Second panel: pulse dispersion as a function of mean
pulse position for the three chains with initial amplitude $p_0=8$.
}
\label{fig2}
\end{figure}

Results for the mean distance traveled by the pulse as a function of time
for isolated chains of 151 sites are shown in the first
panel in Fig.~\ref{fig2}
for the hard, harmonic, and soft potentials
and for various values of the initial pulse amplitude $p_0$. The mean
distance varies essentially linearly with time in all cases (this is only
approximately true in all cases -- even the harmonic oscillator
exhibits early
deviations from linear behavior due to inertial effects).
The important results apparent from Fig.~\ref{fig2} are summarized as
follows and can be understood from the frequency vs energy trends
in Fig.~\ref{fig1}:
\begin{enumerate}
\item
The pulse velocity in the harmonic chain is {\em independent} of the initial
amplitude. This reflects the energy-independence of the mean frequency
(and in fact of the entire frequency spectrum) for harmonic chains
(also see Appendix~B).
\item
The pulse velocity in the hard chain {\em increases} with increasing
initial amplitude.  This is because the mean frequency for the hard chains
increases with increasing energy.
\item
The pulse velocity in the soft chain {\em decreases} with increasing
initial amplitude.  This is because the mean frequency for the soft chains
decreases with increasing energy.
\end{enumerate}
We note that with our choice of potentials the velocity in the hard chain
for very weak initial amplitudes may actually lie below that of the
harmonic chain or even the soft chain because we have omitted a
harmonic contribution to the hard potential, but the
hard chain velocity necessarily increases and surpasses that of the
other chains with increasing initial pulse amplitude.

Not only is the pulse transmitted more rapidly in the hard isolated chains
than in the others, but the pulse retains its integrity over longer
distances
in the hard chain.  This is seen in the second panel in Fig.~\ref{fig2}.
The dispersion $\sigma^2$ is shown for the three chains for a particular
initial pulse amplitude.  Rather than the dispersion as a function of
time, the dispersion is shown as a function of position along the chain so
that the pulse widths at a particular location
along the chain can be compared directly.  Clearly
the hard chain pulse is the most compact {\em at a given distance from the
initially disturbed site} (a plot of $\sigma^2$ {\em vs} $t$ would show the
opposite trend, that is, the pulse in the hard chain would have the
greatest width, but it will have traveled a much greater distance than the
pulses in the other chains).
This combination of results leads
to interesting geometrical consequences in higher dimensions (see
Sec.~\ref{twod}).

\subsection{Chains at Zero Temperature}

If the chains are connected to a heat bath at zero temperature, the equations
of motion Eq.~(\ref{langn}) are modified by the inclusion of the
dissipative contribution,
\begin{equation}
\ddot{x_n} = -\frac{\partial}{\partial x_n} [V(x_n -x_{n-1}) +
V(x_{n+1}-x_n)] -\gamma \dot{x}_n~,
\label{langzerot}
\end{equation}
where $\gamma$ is the dissipation parameter.
The initial conditions are as set forth in Eqs.~(\ref{initialc}).

\begin{figure}[htb]
\begin{center}
\hspace{0.2in}
\epsfxsize = 4.0in
\epsffile{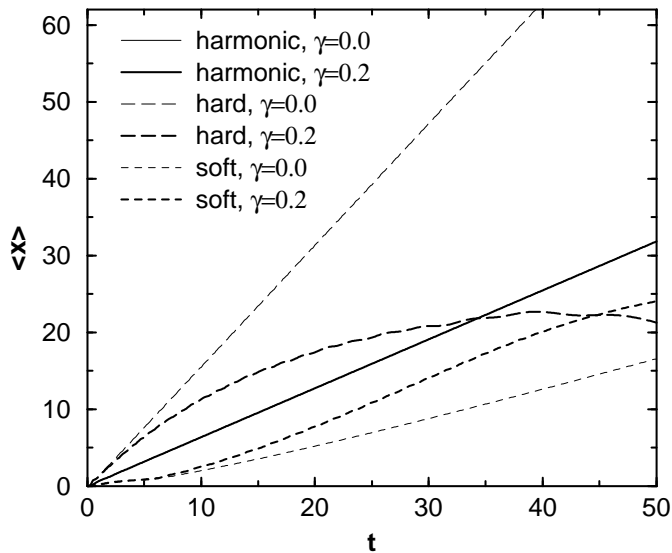}
\vspace{-0.2in}
\end{center}
\caption
{Mean pulse position as a function of time for the
hard, harmonic, and soft chains with initial amplitude
$p_0=8$, with and without friction.}
\label{figsifnot}
\end{figure}

The mean distance traveled by the pulse is shown in Fig.~\ref{figsifnot} for
each of the chains with and without friction so that the frictional effects
can be clearly established.  The salient results can again be understood
from the frequency vs energy trends in Fig.~\ref{fig1}:
\begin{enumerate}
\item
The pulse velocity in the harmonic chain is {\em independent} of friction.
This again reflects the energy-independence of the mean frequency for harmonic
chains. The energy loss suffered through the frictional effects
therefore does not affect the pulse velocity.
\item
The pulse in the hard chain {\em slows down} with time in the
presence of a frictional force.  This is because the chain loses energy via
friction, and the mean chain frequency decreases with decreasing energy.
\item
The pulse in the soft chain {\em speeds up} with time in the presence of a
frictional force.  This is because the chain loses energy via friction, and
the mean chain frequency increases with decreasing energy.
\end{enumerate}

The dependence of the pulse width on friction (not shown explicitly)
follows trends that are consistent with our other results.  An increase in
friction causes the pulse to narrow in the soft chain.  This is consistent
with the observation that higher frequencies are associated with narrower
pulses.  In a harmonic chain there is also some narrowing of the pulse, but
not nearly as much as in the soft chain (detailed explanation of this would
require consideration of the spectrum beyond just the mean frequency).
In the hard chain we can not make
an unequivocal claim from our numerical results because the
dependence of pulse
width on friction for our parameters is extremely weak, with
perhaps a very small amount of narrowing.

\subsection{Chains at Finite Temperature}

If the chains are connected to a heat bath at temperature $T$,
the equations of motion
Eq.~(\ref{langzerot}) are further modified by the inclusion of the
fluctuating contribution,
\begin{equation}
\ddot{x_n} = -\frac{\partial}{\partial x_n} [V(x_n -x_{n-1}) +
V(x_{n+1}-x_n)] -\gamma \dot{x}_n +\eta_n(t)~.
\label{langfinitet}
\end{equation}
The $\eta_n(t)$ are mutually uncorrelated
zero-centered Gaussian
$\delta$-correlated fluctuations that satisfy the fluctuation-dissipation
relation:
\begin{equation}
\langle \eta_n(t) \rangle = 0, \;\;\;\;\;\;\;\;\;\;\;\;
\langle \eta_n(t) \eta_j(t') \rangle = 2 \gamma k_B T
\delta_{nj}\delta(t-t')~.
\label{fdr}
\end{equation}
The initial conditions are now no longer given by
Eqs.~(\ref{initialc}).  Instead, the chain is allowed to equilibrate at
temperature $T$ and at time $t=0$ an additional impulse of amplitude
$\sqrt{2\varepsilon}$ is added to the thermal velocity of site $n=0$.
The integration of the equations of motion proceeds as before, but now
we report averages over 100 realizations. The system is initially
allowed to relax over enough iterations to insure thermal equilibrium,
after which we take our ``measurements."

\begin{figure}[htb]
\begin{center}
\hspace{0.2in}
\epsfxsize = 4.in
\epsffile{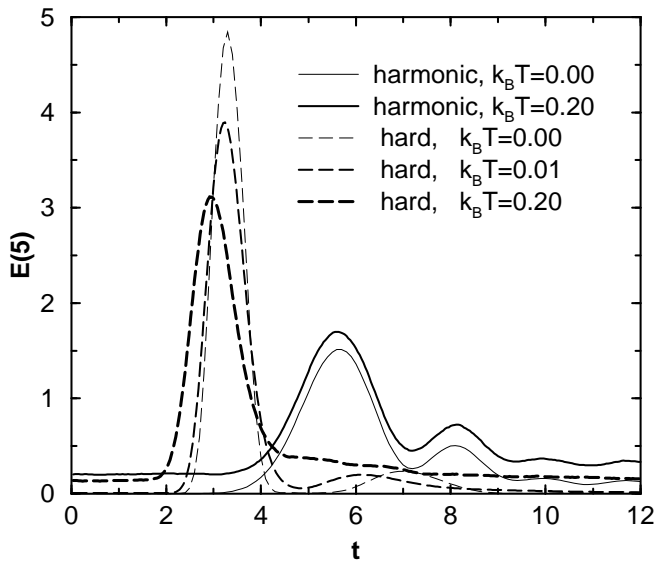}
\vspace{-0.2in}
\end{center}
\caption
{Energy profile vs time at the $5^{th}$ site for hard and harmonic
chains at different temperatures.
The damping parameter in all cases is $\gamma=0.2$ and the initial
pulse amplitude is $p_0=8$.}
\label{figsifsit}
\end{figure}

The pulse dynamics is no longer conveniently characterized by the
mean pulse velocity (although this was the most useful and direct
characterization in the absence of thermal fluctuations).  This is because
there is now a thermal background that causes fluctuations and
distortions of the information in this mean (as well as in other
simple moments and measures such as the pulse maximum).  We find that
the most suggestive presentation of the dynamics is that of the
energy profile itself.  An illustrative set of typical profiles
for chains of 51 sites is presented in Fig.~\ref{figsifsit},
showing energy profiles as a function of time on the $5^{th}$ site on either
side of $n=0$ as a function of temperature.
In all cases there is a delay time until the pulse reaches the
$5^{th}$ site (reflecting a finite velocity). The local energy
around this site then reaches a maximum, and the pulse moves on, leaving
behind a series of later energy oscillations at ever decreasing amplitudes
that eventually settle down to the appropriate thermal levels.  The
after-oscillations are derived analytically in Appendix~\ref{b}
for the harmonic case.  The discussion below
concentrates exclusively on the first pulse, which we think of as
characterizing the arrival of the disturbance at that site.

The important conclusions, some illustrated in the figure, can once
again be understood from the trends in Fig.~\ref{fig1} and
include the following:
\begin{enumerate}
\item
The pulse velocity in the harmonic chain is {\em independent} of
temperature. This is illustrated in the figure by the fact that the peak
of the pulse reaches the particular site under observation at the same time
for the two temperatures shown.  The reason once again is that the
characteristic frequencies of the chain are independent of energy and
therefore the inclusion of thermal effects is immaterial to this measure.
\item
The pulse velocity in the hard chain {\em increases} with increasing
temperature.  This is illustrated by the ever earlier arrival of the pulse
at the site under observation with increasing temperature.  The reason is
that the mean frequency of the chain increases with energy, so that
the hard chain at higher temperatures is associated with a higher frequency
than at lower temperatures and hence with a faster pulse.
\item
The pulse velocity in the soft chain {\em decreases} with increasing
temperature.  This is not explicitly illustrated in the figure, but is
due to the decrease of the mean frequency with increasing temperature.
Thus the soft chain at higher temperatures is associated with a lower
frequency and hence a slower pulse.
\item
The hard chain not only transmits pulses more rapidly than
the other chains, increasingly so with increasing temperature, but it
also transmits the most compact and persistent pulses
at any temperature.  This is seen not only by the obviously smaller
width of the pulses in the hard chain, but by the fact that the
energy trace ``left behind" as the first pulse passes through is
lower in the hard chain than in the other cases.
\end{enumerate}

The pulses in all cases become more dispersive with increasing temperature.
This behavior is clearly evident in Fig.~\ref{figsifsit} for the hard and
harmonic chains, as is the fact that the temperature
dependence of the pulse width is weakest for the hard chain (and strongest
for the soft chain).  These dependences complement those described earlier
for the pulse width as a function of friction: increasing friction in
all cases narrows the pulse (subject to our caveat concerning the
hard chain mentioned earlier) while increasing the temperature broadens it,
both of these dependences being weakest for the hard chain.

\section{Two-Dimensional Isolated Arrays}
\label{twod}

We showed in Sec.~\ref{isolated}
that a pulse travels more rapidly and less dispersively in
an isolated hard chain than in a harmonic or soft chain.  In
higher dimensions these two
tendencies, that of moving faster and that of maintaining the
energy localized, leads to some interesting geometric effects and
to very different pulse propagation properties depending on the
spatial configuration of the initial condition.

In one dimension one could visualize the displacements and momenta
$x,\dot{x}$ as describing motion along the chain or perpendicular
to the chain.  In two dimensions these are distinct cases: a
generalization of the first requires introduction of
two-dimensional coordinates $(x,y)$ and momenta
$(\dot{x},\dot{y})$.  The second requires only a single
perpendicular coordinate $z$ and associated momentum $\dot{z}$ for
each site, and this is the case we pursue.  We thus consider a
two-dimensional square array of dimension $(2N+1)\times(2N+1)$
wherein motion occurs in a direction perpendicular to the array.
The Hamiltonian with $\dot{z}_{n,j}\equiv p_{n,j}$ is expressed as
a sum of local energy contributions,
\begin{equation}
H=\sum_{n,j} E_{n,j}~,
\end{equation}
where
\begin{eqnarray}
E_{n,j}=\frac{p_{n,j}^2}{2}&+&
\frac{k}{4}\left[(z_{n,j}-z_{n+1,j})^2 +(z_{n,j}-z_{n-1,j})^2 +
(z_{n,j}-z_{n,j+1})^2+ (z_{n,j}-z_{n,j-1})^2\right]\nonumber\\
\nonumber\\
&+&
\frac{k'}{8}\left[(z_{n,j}-z_{n+1,j})^4 +(z_{n,j}-z_{n-1,j})^4 +
(z_{n,j}-z_{n,j+1})^4+ (z_{n,j}-z_{n,j-1})^4\right].\nonumber\\
&&
\end{eqnarray}
For the harmonic case we take $k=0.5$ and $k'=0$, and for the hard
anharmonic array we set $k=0$ and $k'=0.5$.  Our lattices are of
size $51\times 51$ and our integration time step is $\Delta
t=0.0005$.  The boundary conditions are immaterial (although we
happen to use boundaries whose edge sites have only three
connections and its corners sites only two) because the lattices
are sufficiently large for the initial excitations not to reach
the boundaries within the time of our computations.

We consider two initial excitation geometries.  In one a ``front" is
created by exciting all sites along the line $(0,j)$, $-N\leq j\leq
N$, with the same initial momentum $p_{0,j}=p_0$.  The front
then moves symmetrically away from this line and its motion is
measured by the mean distance and dispersion
(in all double
sums $n$ and $j$ each range from $-N$ to $N$)
\begin{equation}
\left< x\right>\equiv \frac{\sum_{n,j}
|n| E_{n,j}}{\sum_{n,j}
E_{n,j}}~,
\end{equation}
\begin{equation}
\sigma^2 \equiv \left< x^2\right> - \left< x\right>^2 =
\frac{\sum_{n,j} n^2E_{n,j}}
{\sum_{n,j} E_{n,j}} - \left<
x\right>^2~.
\end{equation}
In the other, an initial pulse of kinetic energy is deposited at the
central site of the array.  We then measure the mean radial
distance of the pulse from the origin,
\begin{equation}
\left< r\right> \equiv \frac{\sum_{n,j}
\sqrt{n^2+j^2} E_{n,j}}{\sum_{n,j}
E_{n,j}}
\end{equation}
(the dispersion in this case is less informative but can also
be monitored if desired).
The motion and dispersion in this geometry are expected to be
roughly spherically symmetric subject to the square connectivity
of the lattice.

\begin{figure}[htb]
\begin{center}
\epsfxsize = 4.0in
\hspace{0.1in}
\epsffile{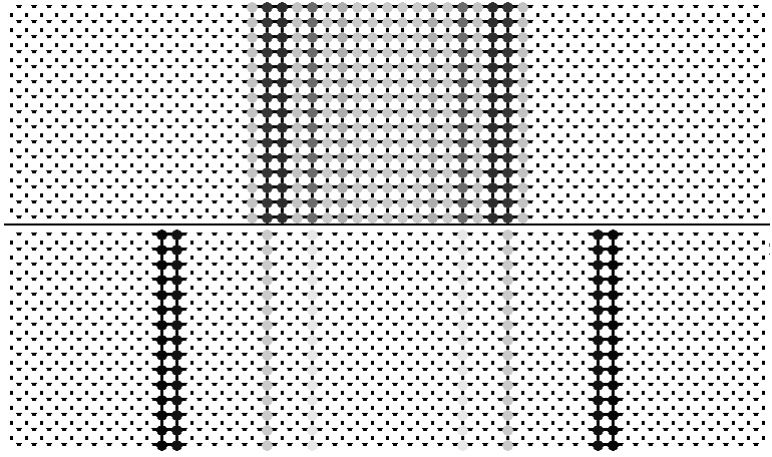}
\end{center}
\caption
{Snapshot at a subsequent time of the energy distribution for
the propagation of an initial front.
Upper panel: harmonic lattice. Lower panel: hard lattice.}
\label{figdibfront}
\end{figure}

\begin{figure}[htb]
\begin{center}
\epsfxsize = 4.0in
\hspace{0.1in}
\epsffile{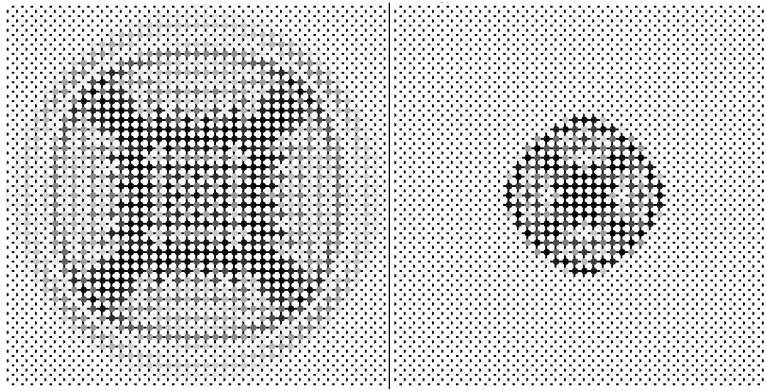}
\end{center}
\caption
{Snapshot at a subsequent time of the energy distribution for
the propagation of an initial pulse at the center of the array.
First panel: harmonic lattice. Second panel: hard lattice.}
\label{figdibpols}
\end{figure}

\begin{figure}[!htb]
\begin{center}
\leavevmode
\epsfxsize = 3.in
\epsffile{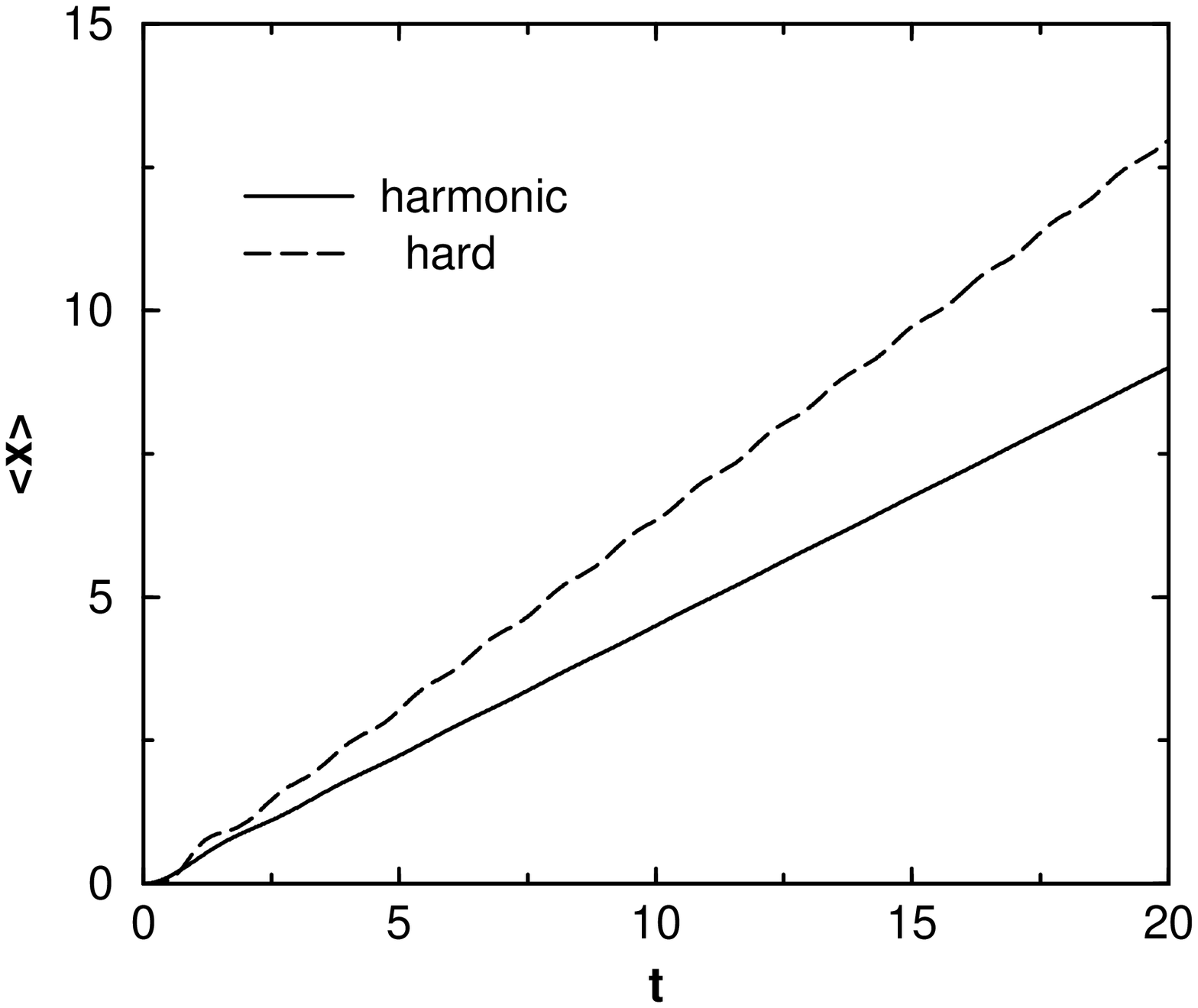}
\leavevmode
\epsfxsize = 3.0in
\epsffile{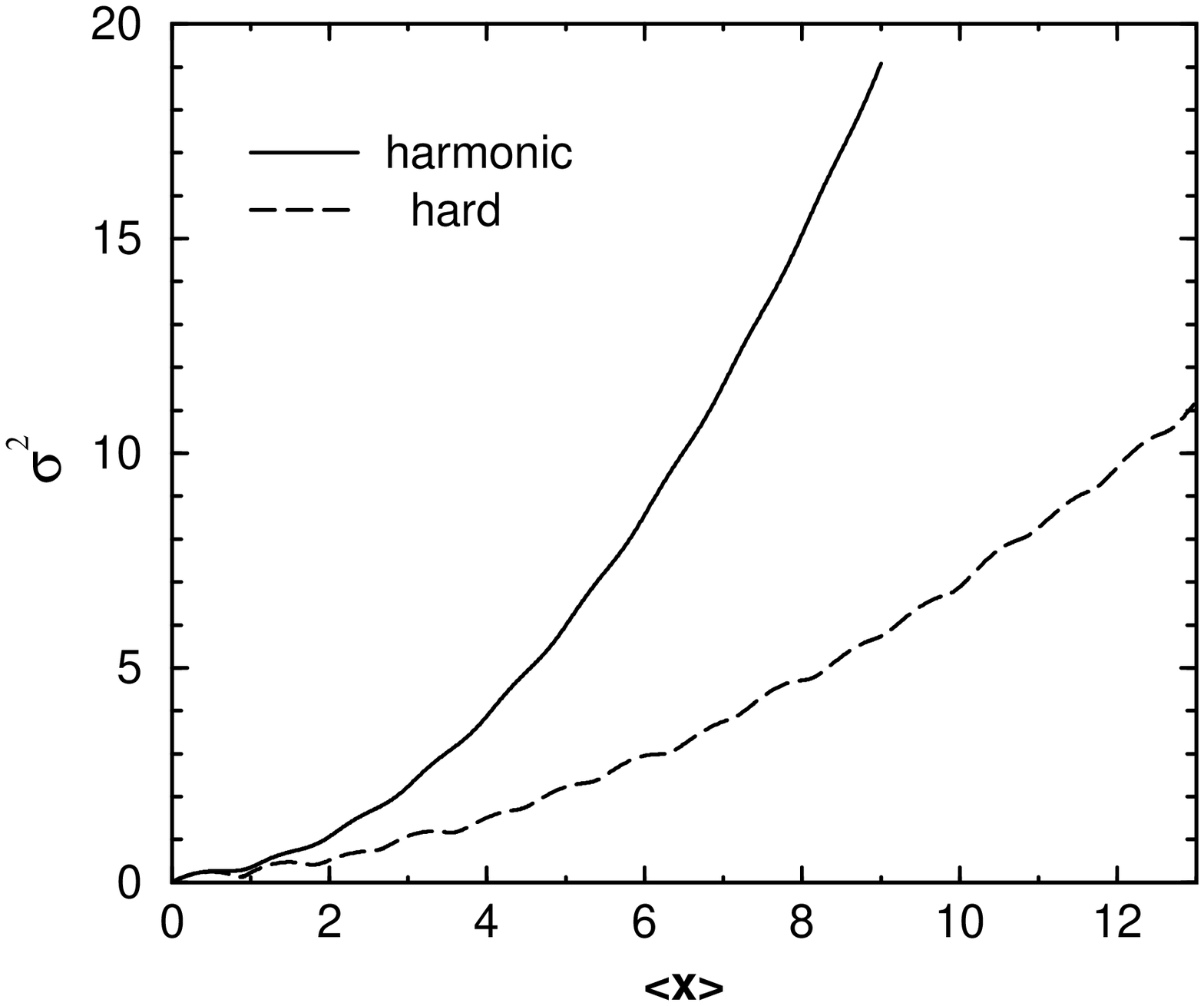}
\vspace{-0.2in}
\end{center}
\caption
{First panel: temporal evolution of the mean distance
$\langle x\rangle$ of
front propagation in harmonic (solid line) and hard anharmonic
(dashed line) lattices.
Second panel: the associated dispersion as a function of the mean distance
traveled.}
\label{figfront}
\end{figure}

\begin{figure}[!htb]
\begin{center}
\hspace{0.2in}
\leavevmode
\epsfxsize = 4.in
\epsffile{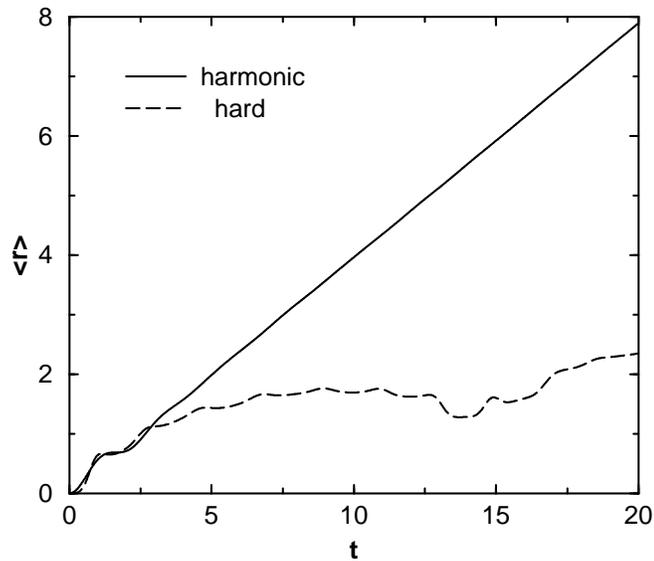}
\vspace{-0.2in}
\end{center}
\caption
{Temporal evolution of the mean radius $\langle r\rangle $ of
pulse propagation
in harmonic (solid line) and hard anharmonic (dashed line) lattices.}
\label{figpols}
\end{figure}

Typical gray-scale snapshots of the energy distribution are shown in
Figs.~\ref{figdibfront} and \ref{figdibpols}, and the differences,
while easily understood, are clearly dramatic.  In the case of the front,
the tendency of a hard lattice to propagate faster than the harmonic
lattice while maintaining the energy more localized is clearly
realized.  The associated mean distance and dispersion that quantify the
comparison are shown in Fig.~\ref{figfront}.  In the case of an initial
point pulse, on the other hand, there is clearly a conflict between
rapid motion and smaller
dispersion -- one can be realized only at the expense of the other.
The latter ``wins": the pulse remains more localized in time
in the hard lattice than in the harmonic.  The associated mean radius
is shown in Fig.~\ref{figpols}.  In the anharmonic lattice the pulse at
first expands as fast as the harmonic but it essentially quickly
saturates while the harmonic pulse continues to disperse.

\section{Conclusions}
\label{conclusions}

In this paper we have considered pulse propagation in discrete arrays
of masses connected by harmonic or anharmonic springs.  We have focused on
the pulse velocity and width, and have found a pattern of behavior that
can be strongly correlated with the energy dependence of the mean
array frequency.

First we investigated the propagation of pulses in isolated
(microcanonical) arrays.  We found that in a hard array
an amplitude increase causes a pulse to travel more rapidly and
less dispersively.
In a harmonic array the pulse speed and width are independent of pulse
amplitude, while in a soft array a more intense pulse travels more slowly
and spreads out more rapidly. These trends are a result of the
fact that in a hard array the mean frequency increases with energy,
in a harmonic array it is independent of energy, and in a soft array the
mean frequency decreases with increasing energy.  In higher dimensions
these trends lead to interesting initial condition dependences that
in turn may lead to apparently ``opposite" behavior in different
cases.  Thus, for
example, a front in a two-dimensional isolated
hard array propagates more rapidly
and more sharply than in harmonic or soft arrays, and the effect is
enhanced if the front is more intense.  On the other hand, a point pulse
in a hard array spreads more slowly than in the others: it is not possible
in this geometry to both propagate quickly and yet retain a strong
localization of energy, and the latter tendency dominates the dynamics.

We then investigated the effects on pulse propagation of connecting the
nonlinear chains to a heat bath (we did this only for the 1D arrays).
We found that dissipative forces tend to slow down the pulse in the hard
array, leave its speed unchanged in the harmonic chain, and actually
speed it up in the soft array.  This somewhat counterintuitive
behavior is, however, fully consistent with the observation that
dissipation causes a decrease in energy and hence a decrease in mean
frequency in the hard case and an increase in mean frequency in the soft
chain (and no change in the mean frequency of the harmonic chain).
Dissipation in all cases causes a narrowing of the pulse, the effect
being greatest in the soft array.

An increase in temperature has the opposite (and again at first sight
perhaps somewhat counterintuitive) effect: it speeds up the pulse in
the hard array, leaves it unchanged in the harmonic array, and slows it
down in the soft chain.  Again this behavior is consistent with the
frequency vs energy trends and the fact that an increase in temperature
is associated with an increase in the energy of the chain.  A temperature
increase in all cases causes a broadening of the pulse, the effect again
being greatest in the soft array.

\section*{Acknowledgments}
R. R. gratefully acknowledges the support of this
research by the Ministerio de Educaci\'{o}n y Cultura through
Postdoctoral Grant No. PF-98-46573147.  A. S. acknowledges sabbatical
support from DGAPA-UNAM.
This work was supported in part by the Engineering Research Program of the
Office of Basic Energy Sciences at the U. S. Department of Energy under
Grant No. DE-FG03-86ER13606.

\appendix

\section{Frequency vs Energy for Oscillator Chains}
\label{a}

Consider a chain of oscillators, and let us focus on the displacement
variable $x$ of a particular mass in the chain, say oscillator j, whose
displacement satisfies the equation of motion
\begin{equation}
\frac{dx_j}{dt} = \pm \left(2 \left[E-\sum_n V(x_n - x_{n-1})
\right]- \sum_{n\neq j} p_n^2\right)^{1/2}~.
\label{oneoscillatorchain}
\end{equation}
The period of oscillation for oscillator j can be defined in analogy with
Eq.~(\ref{omega}):
\begin{eqnarray}
\tau(E;\bf{x'}, \bf{p'}) & = & \frac{2\pi}{\omega(E;
\bf{x'}, \bf{p'}) } \nonumber\\ \nonumber\\
&=&4 \int_0^{x_{max}} \frac{dx_j}{ \left(2 \left[ E-\sum_n V(x_n-x_{n-1})
\right]  -  \sum_{n\neq j} p_n^2\right)^{1/2}}~,\nonumber\\
&&
\label{omegag}
\end{eqnarray}
where
$\bf{x'}$
stands for the set of all the $x$'s {\em except}
$x_j$, and similarly for
$\bf{p'}$.
The upper limit of integration $x_{max}$ depends not only on
$ E$ but on all the
other displacements and momenta, and is the positive value of $ x_j$
at which the denominator of the integrand vanishes.  The resulting
$\omega$ with all the coordinate and momentum dependences is
not very useful, but
it would seem reasonable to simply average over all possible values of these
coordinates and momenta and thus obtain an average period.   We
define the average period as
\begin{eqnarray}
\tau(E)&\equiv& \left< \tau
(E;\bf{x'}, \bf{p'})\right>\nonumber\\ \nonumber\\
& \equiv & 4 \ \frac{
 \int \cdots \int d\bf{x'}\int\cdots \int d\bf{p'}
 \int_0^{x_{max}}dx_j
\left(2\left[E- \sum_ nV(x_n-x_{n-1})\right] -\sum_{n\neq j}
p_n^2\right)}{ \int \cdots \int d\bf{x'} \int \cdots
\int d\bf{p'} }^{-1/2} .\nonumber\\
&&
\label{averagetau}
\end{eqnarray}
The limits of integration not explicitly indicated are appropriate nested
relations among the variables and the energy such that the argument of the
square root always remains positive. The multiple integral in the denominator
covers the same integration regime and insures proper normalization for this
average.  Our interest lies in extracting the energy dependence - the remaining
energy-independent coefficients are complicated and not important for our
arguments. If the pair potentials are powers as in the single
oscillator example, the scaling argument can be generalized by
introducing scaled variables $z_n\sim (x_n-x_{n-1})E^{1/\alpha}$ and
$u_n \sim p_nE^{1/2}$ with appropriate constants of proportionality.
The limits of integration then become independent of energy and
the only energy dependence arises from factoring an $E^{1/2}$ from
the square root in the denominator and an $E^{1/\alpha}$ from
the numerator because it contains one $z-$integration more than the
denominator.
The result, as before, is that
\begin{equation}
\tau(E) = {\mathcal B}_\alpha E^{\frac{1}{\alpha}-\frac{1}{2}}
\end{equation}
with a complicated but {\em energy-independent} expression for the coefficient
$\mathcal{B}_\alpha$, and therefore
\begin{equation}
\omega(E) ~\equiv~ \frac{2\pi}{\tau(E)} \sim
E^{\frac{1}{2}-\frac{1}{\alpha}}~.
\end{equation}

More complicated potentials require suitable generalization of
this argument, but the result in any case is that the average
frequencies for the hard, harmonic, and soft chains follow the
same trends as those shown in Fig.~\ref{fig1}.

\section{Isolated Linear Oscillator Chain}
\label{b}

Although linear oscillator chains are of course fully
understood, it is nevertheless useful to present aspects of their behavior
in the context of the present discussion.

The linear equations of motion (\ref{langn}) with the initial conditions
(\ref{initialc}) are easily solved:
\begin{equation}
x_n(t)  =  {\sqrt{2\varepsilon} \over {2N+1}}  \sum_{q=-N}^N \ {\sin
(\omega_qt) \over \omega_q} \ e^{-2 \pi iqn/(2N+1)}  =  \sqrt{2\varepsilon}
\int_0^t \ J_{2n} (2 \sqrt{k}\tau ) d\tau
\label{displh}
\end{equation}
where the frequencies $\omega_q$ obey the dispersion relation
\begin{equation}
\omega_q^2 = 4k \sin^2\left(\frac{2\pi q}{2N+1}\right)
\label{dispersion}
\end{equation}
and where $J_n(z)$ denotes the Bessel function of the first kind of integer
order $n$. (The energy-independence of the frequencies for the harmonic
chain seen here in the $\varepsilon$-independence of the $\omega_q$
is prominent in our discussions throughout this paper.)
The momenta are then
\begin{equation}
\dot{x_n}(t) = \sqrt{2\varepsilon}\ J_{2n}(2\sqrt{k}t).
\label{velh}
\end{equation}

\begin{figure}[htb]
\begin{center}
\hspace{0.2in}
\epsfxsize = 4.in
\epsffile{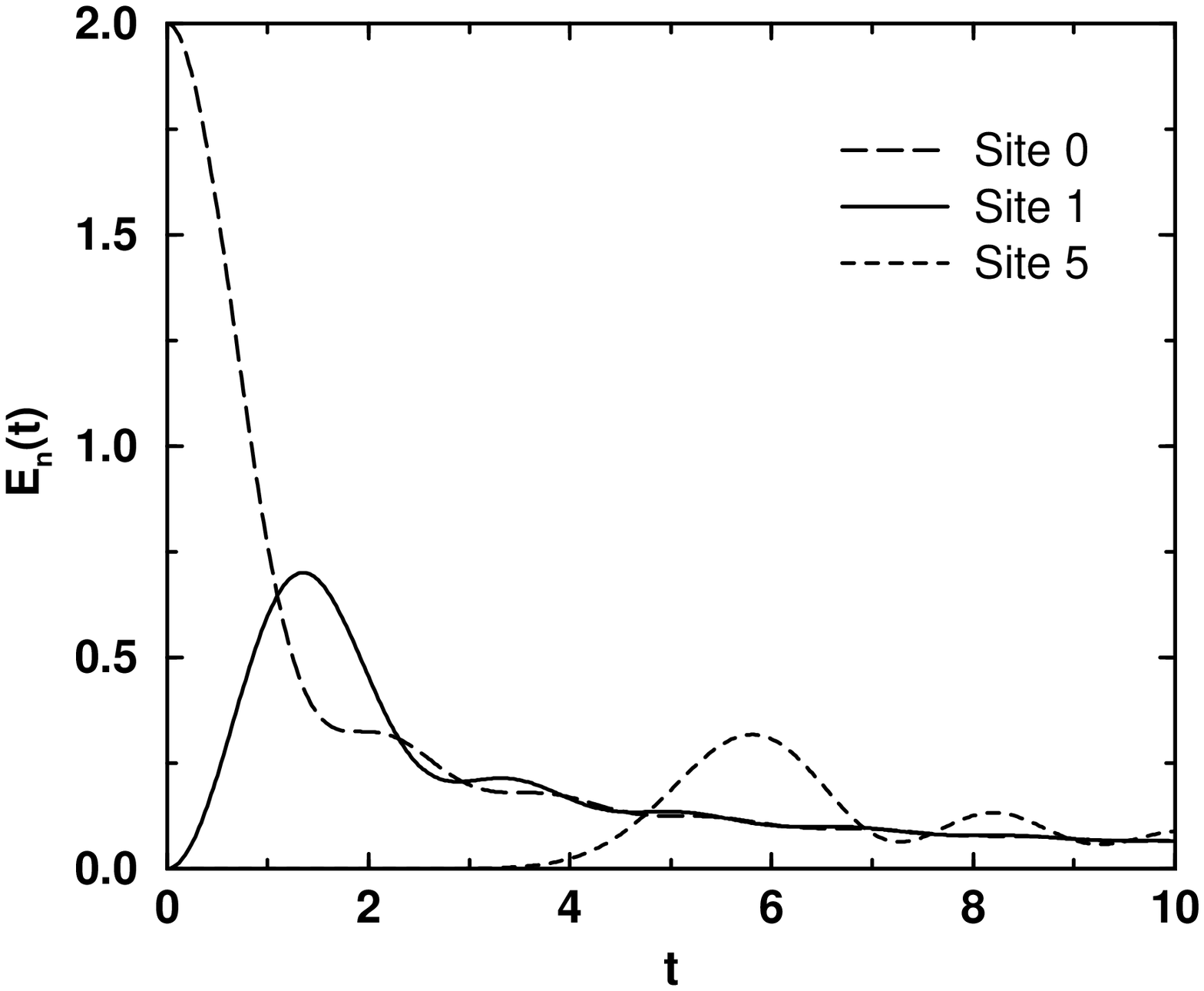}
\vspace{-0.2in}
\end{center}
\caption
{Temporal evolution of the energy $E_n(t)$ at several sites in a harmonic chain
with force constant $k=1$ and initial momentum $p_0=2$.}
\label{figure3}
\end{figure}

Using a number of relations obeyed by the Bessel functions it is
possible to combine these results and obtain for the local energy
the simple expression
\begin{equation}
E_n(t) = \varepsilon \left[ J_{2n}^2(2\sqrt{k} t) +
\frac{1}{2}J_{2n+1}^2(2\sqrt{k} t) +
\frac{1}{2}J_{2n-1}^2(2\sqrt{k} t)\right]~.
\label{harmonicenergy}
\end{equation}
The energy profiles for various sites are shown in
Fig.~\ref{figure3}.  The $n=5$ profile (here obtained from the
analytic expression (\ref{harmonicenergy}) also appears in
Fig.~\ref{figsifsit} (there obtained by numerical integration).
Note that the energy is not transported in a single
absorption-emission process but rather in a series of oscillatory
steps of decreasing amplitude. Our analysis in the body of the
paper focuses on the first energy pulse.

\begin{figure}[htb]
\begin{center}
\leavevmode
\epsfxsize = 3.0in
\epsffile{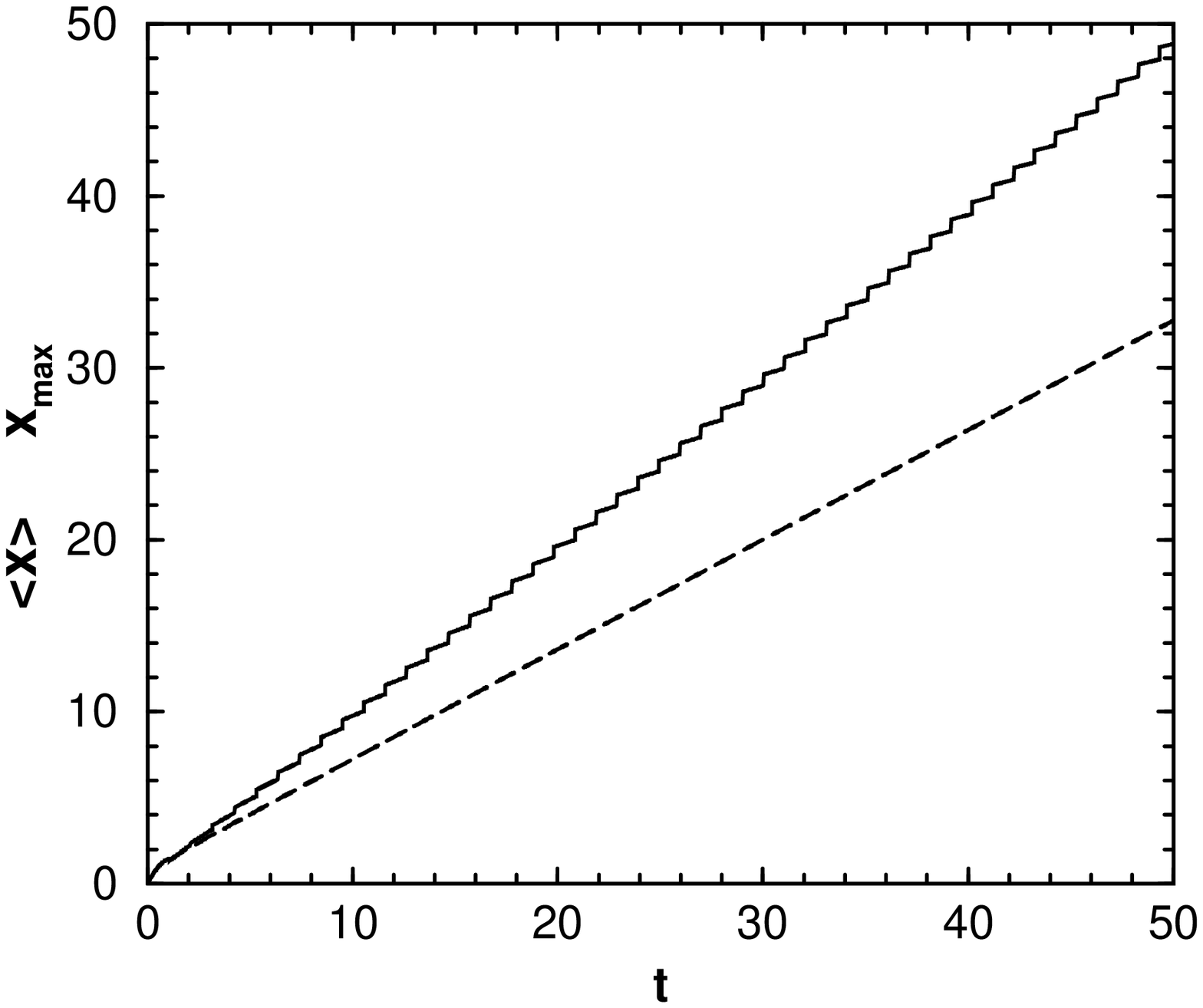}
\leavevmode
\epsfxsize = 3.0in
\epsffile{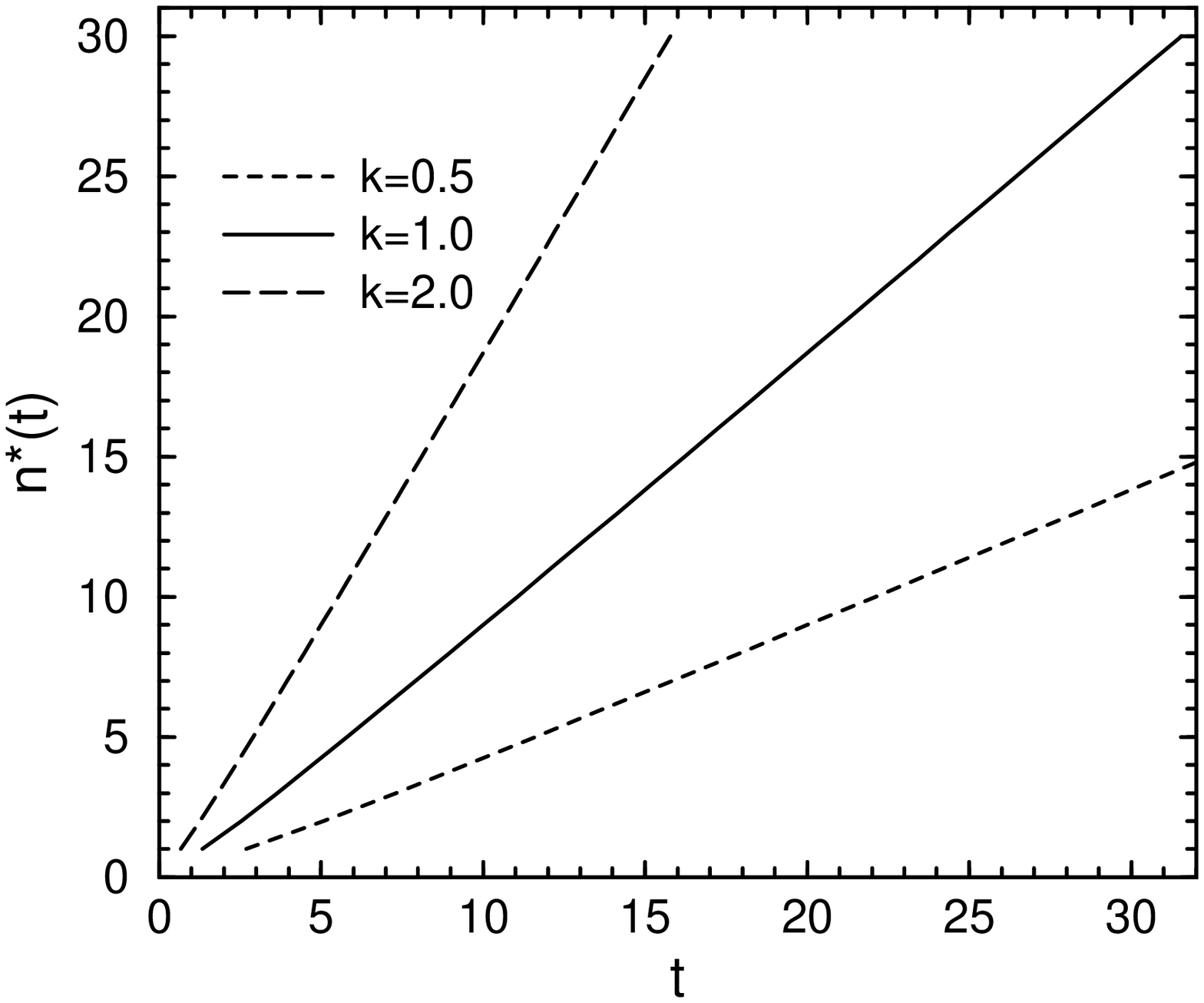}
\vspace{-0.2in}
\end{center}
\caption
{First panel: Mean distance traveled by the pulse (dashed curve) and
pulse maximum (solid curve) as a function of time
for a harmonic chain of unit force constant.
Second panel: Pulse maximum as a function of time for $k=2$
(long-dashed curve), $k=1$ (solid curve), and $k=1/2$ (short-dashed
curve).  The slopes of these numerically generated curves are essentially
as given in the analytic expression Eq.~(\ref{slope}). }
\label{meanvsmax}
\end{figure}

In Section~\ref{isolated} we rely on $\langle x(t)\rangle$,
the mean distance traveled by
the pulse as a function of time, as one measure
to characterize the transport
properties of our arrays.  An alternative measure that can be
calculated analytically for the harmonic chain (but turns out to
be somewhat less convenient for numerical computation) is the
time-dependent site $n^*(t)$ at which the energy is a maximum.
Because the passing energy pulse in general leaves a track behind
it, one expects $n^*(t)\equiv x_{max}(t)$ to grow more rapidly
than $\langle
x(t)\rangle$.  That this is indeed the case is illustrated in
the first panel of Fig.~\ref{meanvsmax}, where both quantities
are shown for a harmonic chain with unit force constant.  The
steps in the $x_{max}$ curve are a  consequence of the
discreteness of the problem.  The analytic result for $n^*(t)$ is
obtained by maximizing Eq.~(\ref{harmonicenergy}) with respect to
$n$ and is, after some manipulation, found to be the solution of
the relation
\begin{equation}
\frac{1+\sqrt{\frac{2n-1}{2n+1}}}{4n} =\frac{J_{2n}(2\sqrt{k}t)}
{2\sqrt{k}tJ_{2n-1}(2\sqrt{k}t)}~.
\end{equation}
Except for a very short initial transient the solution is
essentially linear in time and exceedingly simple:
\begin{equation}
n^*(t)\equiv x_{max}(t)\approx \sqrt{k} t~.
\label{slope}
\end{equation}
This dependence is confirmed in the second panel of
Fig.~\ref{meanvsmax} for three values of the force constant.
The curves shown are obtained numerically, and differ from the
analytic straight lines only at the very earliest times by an
exceedingly small barely visible amount.

\end{document}